# Complementary Fourier single-pixel imaging

Dong Zhou,[1] Jie Cao,[1,*] Huan Cui,[1] Qun Hao,[1] Bing-kun Chen,[1] and Kai Lin[2]

[1] *School of Optics and Photonics, Beijing Institute of Technology, Key Laboratory of Biomimetic Robots and Systems, Ministry of Education, Beijing 100081, China*
[2] *Beijing Fisheries Research Institute, Beijing 100068, China*
*\* ajieanyyn@163.com*

**Abstract:** Single-pixel imaging, with the advantages of a wide spectrum, beyond-visual-field imaging, and robustness to light scattering, has attracted increasing attention in recent years. Fourier single-pixel imaging (FSI) can reconstruct sharp images under sub-Nyquist sampling. However, the conventional FSI has difficulty with balancing the imaging quality and efficiency. To overcome this issue, we proposed a novel approach called complementary Fourier single-pixel imaging (CFSI) to reduce measurements while retaining its robustness. The complementary nature of Fourier patterns based on a four-step phase-shift algorithm is combined with the complementary nature of a digital micromirror device. CFSI only requires two phase-shifted patterns to obtain one Fourier spectral value. Four light intensity values are obtained by load the two patterns, and the spectral value is calculated through differential measurement, which has good robustness to noise. The proposed method is verified by simulations and experiments compared with FSI based on two-, three-, and four-step phase shift algorithms. CFSI performed better than the other methods under the condition that the best imaging quality of CFSI is not reached. The reported technique provides an alternative approach to realize real-time and high-quality imaging.

## 1. Introduction

Conventional optical imaging systems generally use a pixelated array of detectors as the light detection unit. Since the theory of ghost imaging was demonstrated by using quantum-entangled photon pairs in 1995 [1], a bucket detector or a single-pixel detector has gradually been applied in optical imaging systems [2-10]. Single-pixel imaging (SPI), originating from GI, provides scene information by correlating the modulated light patterns generated by a pseudo-thermal source, digital micromirror device (DMD), or other types of spatial light modulators with the intensity of light from the target scene. SPI has the advantages of a wide spectrum, beyond-visual-field imaging and robustness to light scattering and has been adopted in many fields, including 3D imaging [11-14], terahertz imaging [15-17], multispectral imaging [18-20] and scattering medium imaging [21-23]. However, SPI requires a large number of illumination patterns and takes much time [24]. Therefore, achieving a balance between imaging quality and efficiency remains a challenge in the use of SPI.

The current methods for improving the SPI efficiency are mostly concerned with designing different modulate light patterns and applying compressive sensing theory [25-27]. The different types of the modulate light patterns in SPI are random pattern, Hadamard pattern [28], and Fourier pattern [25]. Hadamard and Fourier patterns belong to deterministic orthogonal basis patterns, which are different from non-orthogonal random patterns. Compared with conventional SPI with random patterns, Hadamard single-pixel imaging (HSI) and Fourier single-pixel imaging (FSI) have the advantages of using under-sampled data to reconstruct a sharp image and realize perfect reconstruction in principle. Further, FSI has better imaging efficiency than HSI [27]. In 2015, Zhang et al. [25] presented FSI based on a four-step phase-shift algorithm that can illuminate a scene with phase-shifting sinusoidal-structured light patterns and applies the inverse fast Fourier transform (IFT) algorithm to reconstruct the final image according to the Fourier spectrum. FSI based on a four-step phase-shift algorithm performs the required full sampling measurements which is two times the number of pixels of the illumination pattern. In 2017, Zhang et al. [27] proposed an FSI based on three-step phase-shift algorithm that only requires 1.5 times the number of pixels of the illumination pattern. In 2019, Deng et al. [29] proposed an FSI based on a two-step phase-shift algorithm and demonstrated that it can perform a full sampling measurements equivalent to the number of the pixels of the illumination pattern. However, the three and two-step phase-shift algorithms sacrifice noise robustness to reduce the number of measurements. The differential measurement of the four-step phase-shift algorithm has good noise robustness, but it requires a large number of measurements.

To balance the imaging efficiency and imaging quality, we proposed a novel method based on the complementary nature of DMD and a four-step phase-shift algorithm, that is, complementary Fourier single-pixel imaging (CFSI). DMD can be individually oriented at ±12° with respect to the plane of the array and corresponds to "0" and "1" in the binary pattern. The area of the pattern where "1" locates is naturally complementary to the area where "0" locates. The complementary nature of DMD was proposed by Luo et al. [30], and the signal-to-noise ratio of GI using complementary random patterns and Gerchberg-Saxton-like algorithm increases compared with using conventional random patterns. Ye et al. [31] presented the complementary nature in CGI-based ghost-difference-imaging (GDI) in 2021. They demonstrated multiwavelength-difference GDI and position-difference GDI by Hadamard-based complementary patterns. Multiwavelength-difference GDI can identify more interested object information and position-difference GDI can directly extract edges in a single-round detection. In FSI based on four-step phase-shift algorithm, complementarity exists between patterns of phase-shift 0 and $\pi$, $\pi/2$, and $3\pi/2$. In contrast to conventional FSI, CSFI has high efficiency of FSI based on two-step phase-shift algorithm and noise robustness of FSI based on four-step phase-shift algorithm, which can considerably improve FSI performance.

In this paper, we present the principles of CFSI and demonstrate its advantages using simulations and experiments. We compare the proposed method with the current FSI based on two, three, and four-step phase-shift algorithms and finally summarize the whole work and draw conclusions.

2. **Principles of CFSI**

The principles of CFSI are based on conventional FSI using a four-step phase-shift algorithm. CFSI illuminates the object with phase-shifting sinusoidal-structured light patterns, and the reflected or transmitted light from the object is collected by a single-pixel detector to obtain the Fourier spectrum. The final image of the object is obtained by applying the fast IFT algorithm to the Fourier spectrum.

In FSI based on a four-step phase-shift algorithm, four sinusoid patterns with a phase-shift of $\pi/2$ are used to modulate light. $P^1_\theta$ represents the pattern of phase-shift $\theta$ with the micromirrors of state "1". Fig. 1(a) shows that the two patterns whose phase differs by $\pi$ are complementary, which is similar to the complementary nature of DMD. As shown in Fig. 1(b), a beam of surface light illuminates the plane of the DMD vertically, and micromirrors of state "0" and "1" reflect the light at a deviation of ±24° from the plane normal. Therefore, modulated light with a phase-shift of 0 in the vertical reflection direction of micromirrors of state "1" is obtained, while in the vertical reflection direction of micromirrors of state "0", modulated light with a phase-shift of $\pi$ is obtained. This complementary regular also applies to Fourier patterns with a phase-shift of $\pi/2$ and $\pi 3/2$. The illumination pattern $P^d_\theta$ is expressed as follows:

$$P^d_\theta\left(x,y;f_x,f_y\right) = a + b\cos\left(2\pi f_x x + 2\pi f_y y + \theta\right), \tag{1}$$

where $a$ is a DC term and represents the average intensity of the image; $b$ represents the contrast; $f_x$ and $f_y$ represent the spatial frequency; $x$ and $y$ represent the 2D Cartesian coordinates in the scene; $\theta$ represents the phase parameter; and $d$ is 0 or 1 depending on the state of "0" or "1" of DMD. The patterns are shown in Fig. 1(c), and the patterns of $P^1_\theta$ or $P^0_\theta$ should be generated and illuminated. Therefore, the values of $\theta$ are equal to 0 (or $\pi$) and $\pi/2$ (or $3/2\pi$), respectively. The number of patterns for full sampling required by CFSI is half of the four-step phase-shift FSI, two thirds of the three-step phase-shift FSI, and equal to the two-step phase-shift FSI. The obtained intensity values after illuminating the patterns on the object are expressed as

$$I^d_\theta(f_x, f_y) = \iint_\Omega O(x, y) P^d_\theta(x, y; f_x, f_y) dx dy, \qquad (2)$$

where $\Omega$ is the illumination area and $O(x, y)$ is the object to be imaged. The Fourier spectrum of $O(x, y)$ can be obtained by processing $I^1_0$, $I^0_0$, $I^1_{\pi/2}$, and $I^0_{\pi/2}$. The formula is expressed as

$$F\{O(x,y)\} = \frac{1}{2b}\left[I^1_0(f_x, f_y) - I^0_0(f_x, f_y)\right] + j \cdot \left[I^1_{\pi/2}(f_x, f_y) - I^0_{\pi/2}(f_x, f_y)\right], \qquad (3)$$

where $F\{O(x,y)\}$ is the Fourier spectrum, and $j$ is the imaginary unit. Although two-phase-shifted patterns are generated for illumination, Eq. (3) also realizes differential measurement in a symmetric manner for removing $a$. Compared with FSI based on a two- or three-step phase-shift, the proposed method can effectively eliminate the noise caused by environmental illumination. Finally, the final image $O'(x, y)$ is obtained by IFT. The formula is expressed as

$$O'(x,y) = \frac{1}{2b}F^{-1}\left\{\left[I^1_0(f_x, f_y) - I^0_0(f_x, f_y)\right] + j \cdot \left[I^1_{\pi/2}(f_x, f_y) - I^0_{\pi/2}(f_x, f_y)\right]\right\}, \qquad (4)$$

where $F^{-1}$ represents the IFT operator.

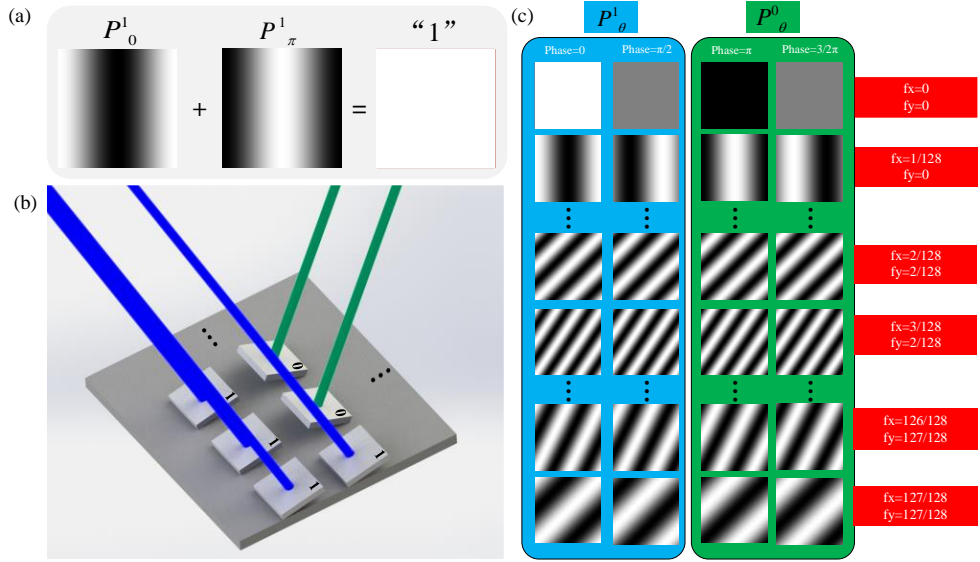

Fig. 1. Principles of CFSI. (a) Two patterns whose phase differs by $\pi$ are complementary. (b) Schematic diagram of generating complementary illumination. Two modulated reflected light whose phase-shifts differ by $\pi$ can be obtained when DMD is loaded with one preset pattern. (c) Complementary Fourier patterns. Only $P^1_\theta$ is loaded to DMD, and the light intensity from the target scene of $P^0_\theta$ can be obtained based on complementary nature.

Natural Fourier basis patterns are grayscale and cannot be directly loaded into the DMD. In previous research, two typical solutions are used to binarize the grayscale Fourier patterns. One is spatial dithering strategy, and the other is temporal dithering strategy [24,32]. As shown in

Fig. 2(b), the patterns of temporal dithering strategy are successive binary patterns that are decomposed from the grayscale Fourier pattern denoted as B0, B1, …B7. These successive binary patterns are illuminated for different times, and patterns of higher order take more time to illuminate. This temporal dithering strategy is used to approximately replace a grayscale pattern with eight binary patterns. Although grayscale FSI is realized by temporal dithering strategy, it requires a large amount of illumination and takes much time for sampling. Fig. 2(c) shows that spatial dithering can accelerate the speed of image acquisition of FSI at the expense of the imaging spatial resolution. A grayscale pattern is replaced by a binary pattern processed by spatial dithering strategy. FSI based on spatial dithering strategy is binary FSI. In this paper, we used binary and grayscale FSI to verify CFSI performance through simulations and used binary FSI to verify CFSI through experiments.

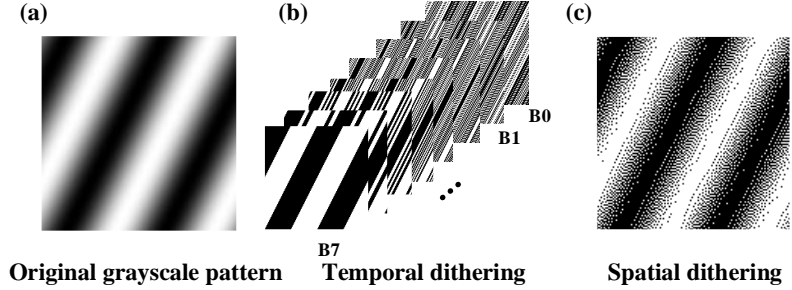

Fig. 2. Methods to binarize the grayscale Fourier patterns. (a) Original grayscale Fourier pattern, (b) binary Fourier pattern binarized by temporal dithering, and (c) binary Fourier pattern binarized by spatial dithering.

In binary CFSI, we also theoretically analyzed the feasibility of complementary patterns instead of patterns with a phase-shift difference of π. Fig. 3 shows that the complementary patterns are similar to the patterns with a phase-shift difference of π but are not exactly the same. Thus, CFSI is not exactly a two-step transformation of FSI based on a four-step phase-shift algorithm.

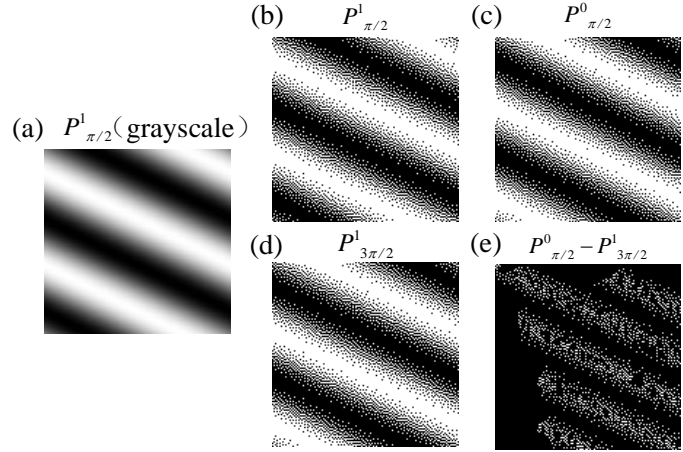

Fig. 3. Comparison of complementary binary pattern and phase-shift difference π binary pattern. (a) Grayscale original Fourier basis pattern with phase-shift π/2 ($f_x=1/128$, $f_y=126/128$). (b) Binary Fourier basis pattern binarized from the original $P^1_{\pi/2}$ with spatial dithering algorithm. (c) Complementary pattern of binary $P^1_{\pi/2}$. (d) Phase-shift difference π pattern of binary $P^1_{\pi/2}$. (e) Difference of complementary binary pattern and phase-shift difference π binary pattern.

## 3. Simulations and experiments

*3.1 Simulations*

Simulations using grayscale and binary patterns were performed separately to evaluate the performance of CFSI compared with FSI based on two-, three-, and four-step phase-shift algorithms.

The performance of the final reconstructed image was compared quantitatively using the peak signal-to-noise ratio (PSNR) [33] as the evaluation index

$$\begin{cases} \text{PSNR} = 10\log_{10}\dfrac{\left(2^k-1\right)^2}{\text{MSE}} \\ \text{MSE} = \dfrac{1}{M}\sum_{x,y}\left(O'(x,y)-O(x,y)\right)^2 \end{cases}, \tag{5}$$

where *MSE* is the mean square error, *M* is the number of the pixels of the whole image, and *k* is the number of bits set as 8. PSNR can reflect the imaging quality; the higher the PSNR, the better the imaging quality.

**a) grayscale patterns**

The reconstruction was simulated using the tested image "cameraman" whose size is 128×128 pixels. We compared four different methods under the measurements 600, 1200, 1800, 2400, 3000, 3600, 6400, and 16384. For CFSI and a two-step phase-shift FSI, full sampling was achieved when the measurements reached 16384. Figs. 4 and 5 show that the imaging quality improves with the increase in measurements. The PSNR of CFSI was higher than that of the three other methods, and the two-step phase-shift FSI was slightly better than the three-step phase-shift FSI. Conventional FSI based a on four-step phase-shift has the worst performance for requiring four illuminations to achieve one spectrum value.

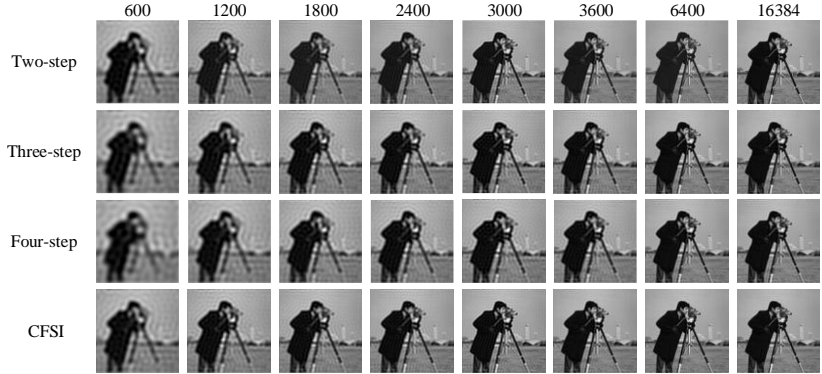

Fig. 4. Reconstructed images of grayscale CFSI and two-, three- and four-step phase-shift grayscale FSI under different measurements.

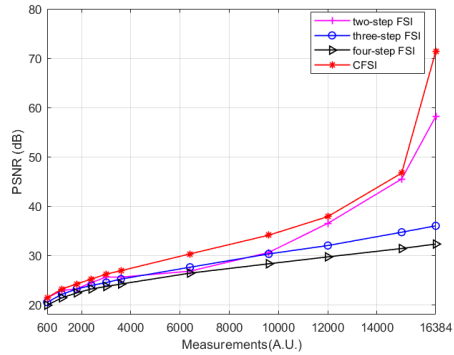

Fig. 5. PSNR values of grayscale CFSI and two-, three-, and four-step phase-shift grayscale FSI under different measurements.

The above results were obtained under the condition of no noise. However, in the actual application environment, noise, such as ambient light and circuit current, is usually introduced into the imaging system. Therefore, we performed another simulation and added measurement noise to evaluate the performance of the four methods. The measurement noise set was Gaussian white noise shown as

$$N(c) = \frac{1}{\sqrt{2\pi}\sigma} exp\left(-\frac{c-\mu^2}{2\sigma^2}\right), \qquad (6)$$

where $c$ is the noise, $\mu$ is the average value, and $\sigma$ is the standard deviation. We set $\mu$ as 0 and set $\sigma$ as 0, 0.1, 0.5, 1, 3, 5, and 10. Measurements were set as 3000. The corresponding reconstructed results are shown in Fig. 5, indicating that the imaging quality of CFSI is better than that of the three other methods under the condition of existing noise, except when the noise level is high enough. The FSI based on a two-step phase-shift is most affected by noise because it is not a differential measurement. The PSNR difference of CFSI is higher than that of the four-step phase-shift FSI because CFSI doubles the spectrum value and introduces twice the noise, which is 1.5 times compared with the three-step phase-shift FSI.

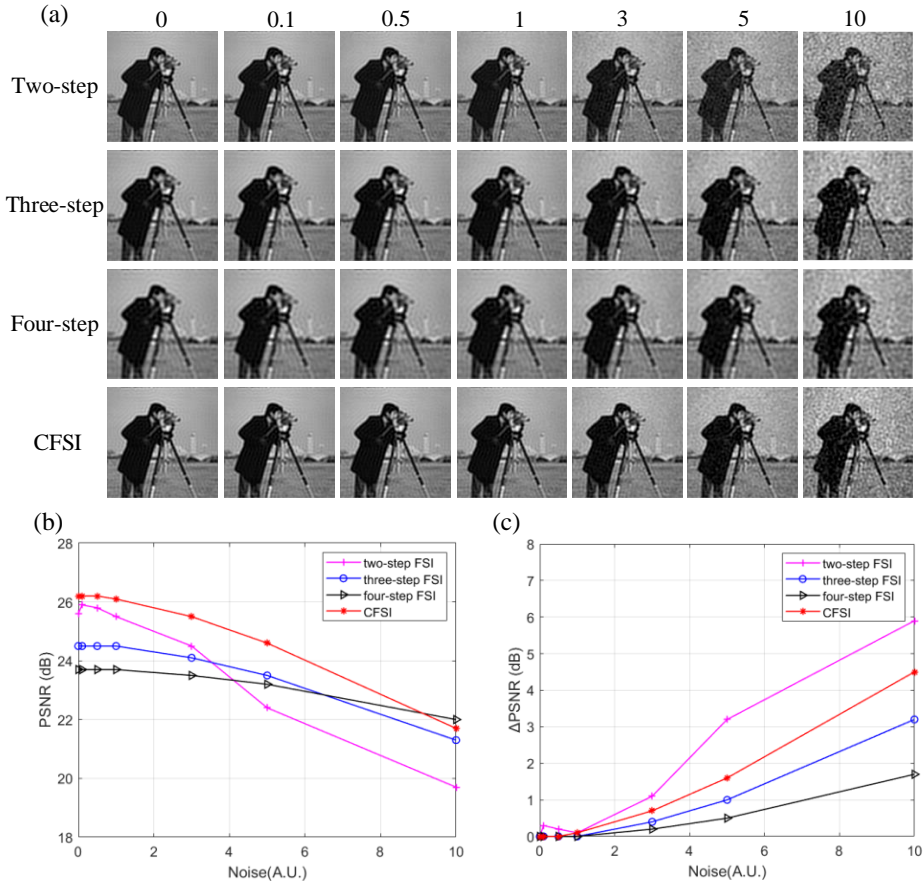

Fig. 6. Reconstructed results of grayscale CFSI and two-, three-, and four-step phase-shift FSI under different standard deviation noise. (a) Final images of the four methods under different standard deviation noise. (b) PSNR values of the final images. (c) PSNR difference between final images and noiseless images.

**b) binary patterns**

We also performed simulations on FSI with binary patterns based on spatial dithering strategy. The setup was the same as that of applying grayscale patterns. The reconstructed results without noise are shown in Figs. 7 and 8, indicating that the results are different from those obtained by applying grayscale patterns. Given the quantization errors and reduction in the imaging spatial resolution caused by spatial dithering strategy [32], the imaging quality is worse than that of the grayscale FSI. The imaging quality of the two-step phase-shift FSI is worse than that of the three other methods because its base pattern is not applicable after spatial dithering [29]. Qualitative comparison indicates that as the measurements increase, the PSNR values of binary CFSI and three- and four-step phase-shift binary FSI increases first and then decreases. CFSI is the first to achieve the best imaging quality, and before that, the PSNR of CFSI is better than the three- and four-step phase-shift binary FSI.

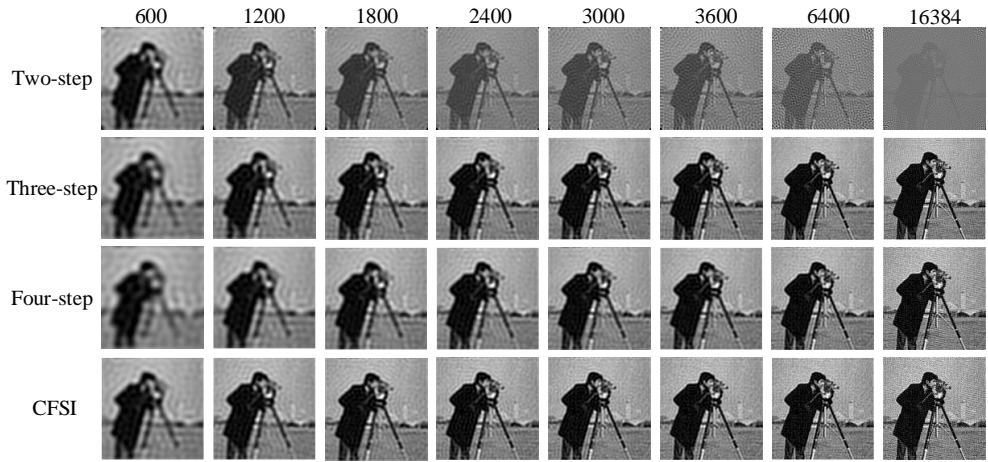

Fig. 7. Reconstructed images of binary CFSI and two-, three-, and four-step phase-shift binary FSI under different measurements.

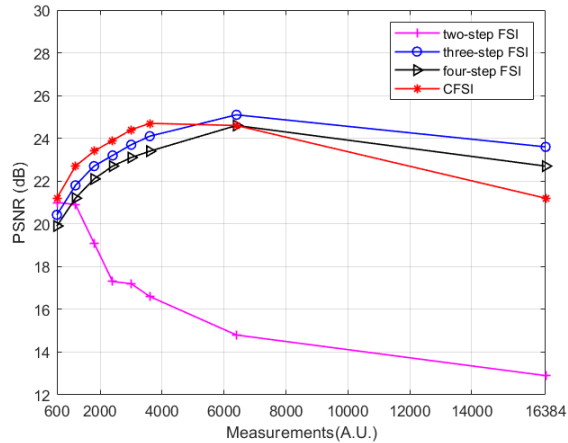

Fig. 8. PSNR values of binary CFSI and two-, three-, and four-step phase-shift binary FSI under different measurements.

The reconstructed results with noise are shown in Fig. 9. Given the poor imaging quality of the two-step phase-shift FSI, the interference caused by noise is lower than that when grayscale patterns were applied; thus, the PSNR difference becomes the smallest. Although the PSNR difference of CFSI is larger than that of the three- and four-step phase-shift FSI, the PSNR value of CFSI is higher than the two other methods under the condition that standard deviation of noise is below 5.

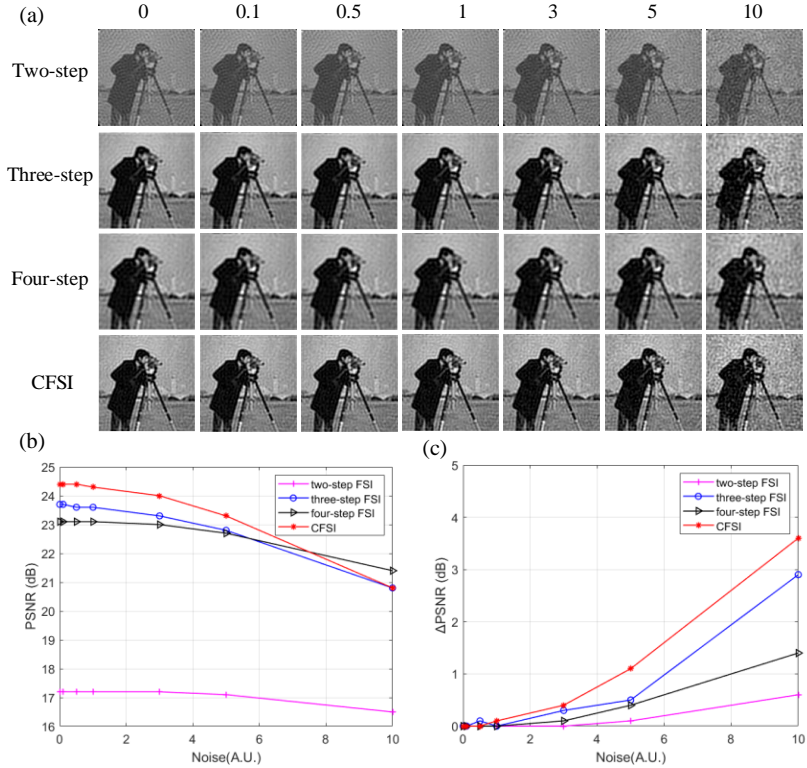

Fig. 9. Reconstructed results of binary CFSI and two-, three-, and four-step phase-shift binary FSI under different standard deviation noise. (a) Reconstructed images of the four methods under different standard deviation noise, (b) PSNR values of the final images, and (c) PSNR difference between the final images and noiseless images.

*3.2* Experiments

The experimental setup is shown in Fig. 10. The setup consists of three parts, namely, the illumination system, the detection system, and the object. The illumination system consists of a light-emitting diode operating at 400–760 nm (@20W), DMD (Texas Instruments DLP Discovery 4100 development kit), and lens. The DMD of the DLP development kit consists of 1024×768 micromirrors, and its maximum binary modulation rate is up to 22 KHz. The focal length of the lens is 150 mm. The detection system consists of two photodetectors (Thorlabs PDA36A, active area of 13 $mm^2$) and collecting lens of focal length 5 mm, data acquisition board (Gage CSEG8, sampling at 1 MS/s), and a computer.

Grayscale FSI with temporal dithering strategy will considerably prolong the sampling time and consume much storage space of DMD. Besides, environmental noise is inevitable in the experimental environment. Although we carried out the experiments in a confined space, when the light illuminated on the target surface and the other blank area, the reflected light fills the entire experimental space due to the large divergence angle of white light. Thus, we performed experiments by adding ambient light noise from light source without adding extra noise. The level of ambient light noise depends on the working power of the light source. During our whole experiments, the light source works at the maximum power. Before the experimental results are shown, some settings should be illustrated first. The resolution of DMD is 1024×768, which is not suitable for applying 128×128 pixels binary patterns if all micromirrors are used. We combined 6×6 micromirrors as one cell, and the patterns occupied part of DMD with 768×768 resolution. In FSI based on two-, three-, and four-step phase-shift algorithms, patterns with the remaining part of DMD set as 0 can be used for sampling. However, in CFSI, the 768×768 pixel binary Fourier patterns were placed at the middle of the DMD, and the remaining two 768×128 parts on both sides were occupied by random patterns. As such, the interference of the unused part of the DMD can be eliminated through differential CFSI measurement.

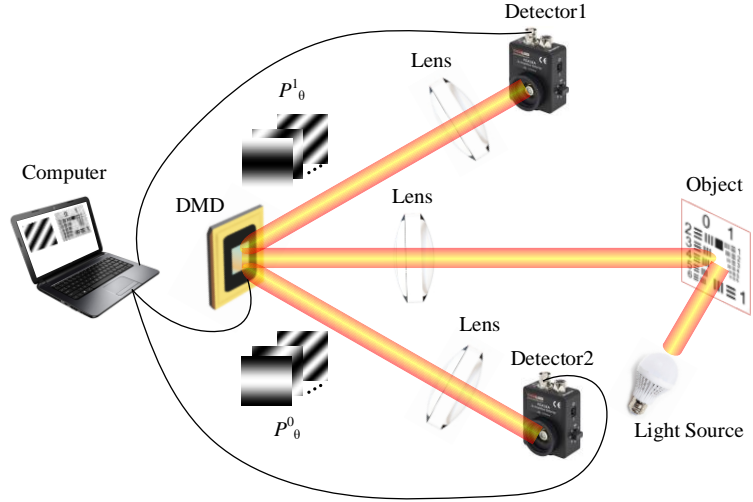

Fig. 10. Experimental setup.

We used a modified US Air Force resolution test chart for imaging. Measurements were set as 600, 1200, 1800, 2400, 3000, 3600, and 6400.

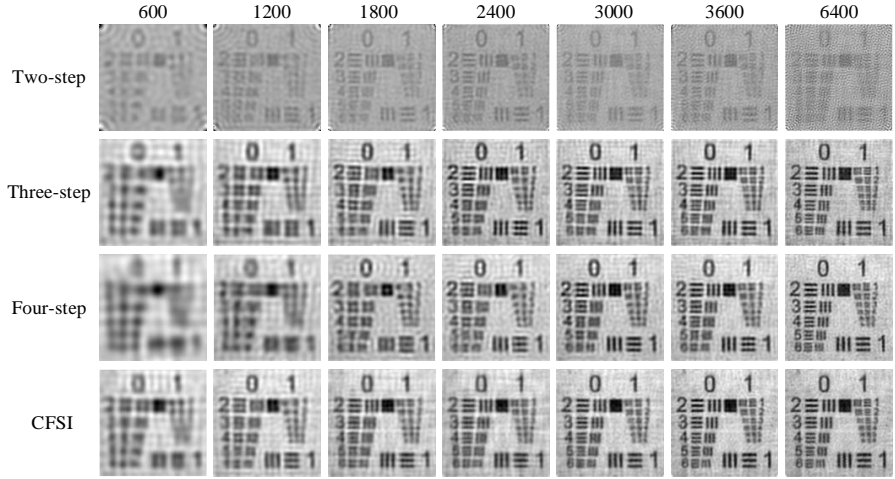

Fig. 11. Experimental results of CFSI and two-, three-, and four-step phase-shift FSI under different measurements.

The experimental results are shown in Figs. 11 and 12, indicating that the imaging quality of the three- and four-step phase-shift FSI improved as the measurements increased. The imaging quality of CFSI and the two-step phase-shift FSI increased first and then decreased, which is similar to the simulation using binary patterns. This reason is that FSI using binary patterns after spatial dithering strategy is not able to realize perfect reconstruction [32]. After reaching the best image quality that can be reconstructed by binary patterns, increasing the number of measurements will decrease the imaging quality. CFSI was the first to achieve its best imaging quality, and before that, the PSNR values of CFSI were higher than those of the three other methods.

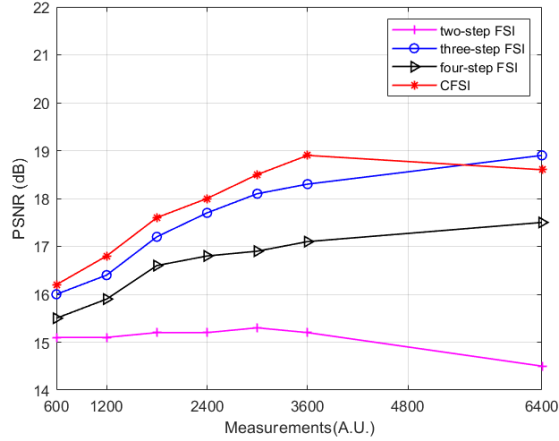

Fig. 12. PSNR values of CFSI and two-, three-, and four-step phase-shift FSI under different measurements.

## 4. Discussion and conclusions

According to the complementary nature of Fourier patterns based on a four-step phase-shift algorithm and the complementary nature of DMD, CFSI can improve the performance of FSI in terms of noise robustness and imaging efficiency. Conventional FSI does not balance imaging efficiency and imaging quality well. CFSI is a differential measurement based on a four-step phase-shift FSI and has the same imaging efficiency as the two-step phase-shift FSI. The results of simulations and experiments suggest that CFSI has the best imaging quality among the three other methods in an environment with noise under the condition that the best imaging quality of CFSI is not reached. However, CFSI also has several disadvantages. First, complementary patterns after Floyd–Steinberg dithering do not perfectly replace binary patterns with a phase-shift difference of $\pi$ [32]. Using Floyd–Steinberg dithering has a certain effect on the imaging quality of CFSI when the four-step phase-shift formula is used for reconstruction. Second, the complementary region of DMD will introduce unused micromirrors that may cause extra intensity value. In our work, we propose that the unused region should be filled with symmetrical random patterns. Third, in our experiment, we use two detectors to obtain the light intensity value of the complementary patterns. The increase of system complexity is inevitable for collecting the reflected light from different directions of the DMD. But the experimental setup can still be improved to use one detector to collect both the light of complementary patterns, and just need a large enough optical path difference to distinguish the two intensity value signals of complementary patterns.

The performance of SPI is mainly affected by two aspects, one is sampling and the other is reconstruction. In terms of sampling, although there are many compressed sensing technologies applied to improve the imaging efficiency of SPI [15,16,34-38], the time-consuming reconstruction process limits their application in real-time imaging [27]. Conventional FSI has the advantage of simple reconstruction for only one IFT is required, and it has been applied in real-time and high-quality imaging [29,32]. The reported method CFSI combines the advantages of current methods and improves the performance of conventional FSI. Therefore, CFSI provides an alternative approach to realize real-time and high-quality imaging.